\documentclass[twoside]{JINST}
\usepackage{graphicx}
\usepackage{rangecite}
\usepackage{amsmath}
\usepackage[figuresright]{rotating}
\usepackage{afterpage}

\title{Pulse-shape discrimination of scintillation from alpha and beta particles with liquid scintillator and Geiger-mode multipixel avalanche diodes}

 \author{I.~Kreslo\thanks{Corresponding author.}, I.~Badhrees, S.~Delaquis, A.~Ereditato, S.~Janos, M.~Messina, U.~Moser, B.~Rossi, M.~Zeller \\
 \llap{}Albert Einstein Center for Fundamental Physics, \\
 Laboratory for High Energy Physics, \\
 University of Bern, \\
 Sidlerstrasse~5,\\
 3012 Bern, Switzerland \\
}

\abstract{A successfull application of Geiger-mode multipixel avalanche diodes (GMAPDs) for pulse-shape discrimination in
alpha-beta spectrometry using organic liquid scintillator is described in this paper. Efficient discrimination
of alpha and beta components in the emission of radioactive isotopes is achieved for alpha energies above 0.3~MeV. The ultra-compact design 
of the scintillating detector helps to efficiently suppress cosmic-ray and ambient radiation background.  This approach allows construction of hand-held robust devices for monitoring of radioactive contamination in various environmental conditions.}


\begin{document}

\section{Introduction}

The technique of distinguishing neutron, alpha and beta activity of radioactive isotopes by pulse-shape discrimination (PSD) in
organic liquid scintillators has been developed since the late fifties of the last century \cite{Brooks,Roush}. The principle of discrimination is based on the different balance between two main mechanisms of de-excitation of the scintillating medium for low ionizing particles (electrons) and highly ionizing particles (alpha). Passage of charged particles through a scintillating medium (usually aromatic composits) leads to excitation and ionization of its molecules. Both singlet and triplet excited molecular states are produced. However, mainly singlet states may lead to the emission of scintillation light. Photon emission from excited triplet state is suppressed by spin conservation. Excited singlet states quickly emit photons, producing prompt scintillation pulses. The time distribution of the light intensity in this pulses follows an exponential decay law with a characteristic decay time $\tau$ of the order of nanoseconds for aromatic liquid scintillators:

$$I_p(t)=I_{0p} e^{-t/\tau}.$$

Triplet states have longer life times (up to seconds). Usually they release excitation by radiationless dissociation. However, there is a finite probability for two excited triplet states to interact and form two singlet states with at least one of them excited. Such a singlet state again releases its excitation by emitting a photon. This emission is delayed with respect to the prompt scintillation, forming a long decay component of the scintillation pulse. The time distribution of the delayed emission is defined by the diffusion of the triplet states:

$$I_d(t)=\frac{I_{0d}}{[1+A ln(1+t/t_a)]^2(1+t/t_a)},$$ 

where $A$ is a constant depending of the scintillating medium, and $t_a$ is the relaxation time defined by the diameter of the ionized column along the particle track $r_0$ and diffusion coefficient of triplet states $D_T$ as $t_a=r_0^2/4D_T$.
This mechanism was first proposed by Parker and Hatchard in \cite{Parker,Parker2} and further developed by Laustriat \cite{Laustriat} to explain differences in scintillation pulse shapes for alpha and beta particles in liquid scintillators. The intensity $I_{0d}$ of the long decay component is defined by the rate of triplet-triplet exchanges, hence by the concentration of excited triplet states.
This concentration, in turn, depends on the Linear Energy Transfer (LET, the amount of energy deposited by the charge particle per unit length of the track). In the energy range of 0.1-10 MeV the LET for heavy alpha particles is orders of magnitude higher than for light electrons. Therefore, the long decay component is stronger for alpha particles than for electrons. 

Gamma radiation in the above energy range reveals itself by the production of Compton electrons and photoelectrons. Hence, the scintillation pulse shape from gamma photon interactions is identical to the one of beta particles. Neutrons in the above energy range interact with the organic medium mostly by elastic scattering, producing recoil protons. These protons have typical LET in between those of alpha and beta particles. Analysis of the time distribution of the scintillation light is the basis for the PSD method for identification of neutron, alpha and beta-gamma radiations.

The total fraction of energy converted into scintillation strongly depends on the particle LET. Due to the interactions of excited singlet states with each other, with ionized molecules and ionization electrons, their excitation can be released in a non-radiative way into ionization and dissociation. This effect is called self-quenching and is described in detail in \cite{Birks}. The light yield per unit length of  particle tracks in organic scintillators is described by the well known semiempiric Birk's formula $dL/dx=A(dE/dx)/[1+k_B(dE/dx)]$, where $dE/dx$ is the particle LET, $A$ and $k_B$ are empirically measured constants.  In organic scintillators, because of this effect, the scintillation light yield per MeV of deposited energy from beta particles (low LET) is about 10 times higher than for alpha particles.

Liquid organic scintillators are widely used
in combination with PSD to quantify and, to a certain extent, to qualify the presence of radioactive contaminants in
different objects, such as water, air, surfaces, soil, etc. The advantage of liquid scintillators is that they provide
very close contact of the scintillating medium with the emitting radioactive atoms. This is especially important
for detecting alpha radiation, since the range of alpha particles from typical radioactive isotopes is very short in dense media,
of the order of 10~$\mu$m. The most successful technique is based on so-called extractive scintillating
compounds, which are capable of chemically dissolving the isotope conglomerates into a homogeneous liquid,
where the radioactive atoms are present in isolated form surrounded by the scintillating molecules. Such an approach
leads to a relatively high energy resolution of the resulting detector, close to the theoretical limit for liquid scintillator.

The scintillating light is typically detected by a vacuum photomultiplier tube (PMT). The size of the tube and the related electronics 
usually defines the dimensions of the instrument. Existing designs have a form of a table-top apparatus with a crate for 
data acquisition and analysis system. This allows for only a limited use in field applications. The volume of the scintillating
cell is usually designed to match the input window of the PMT. This volume defines the rate of the light pulses related to the cosmic rays and the ambient
radiation. This rate limits the sensitivity of the detector to the amount of radioactive contaminants and must be minimized.
Different techniques are used to suppress this background, such as passive screening of the spectrometer cell with low-radioactive materials,
or surrounding it with scintillating detectors to provide a veto signal for external incoming particles. The second method, however, is only efficient to suppress background from high energy electrons and muons, but not from ambient gamma-rays.

The detector presented in this paper has the advantage of being ultra-compact, hence allowing to diminish the rate of background events.
The active volume of the scintillator is of the order of only 1~mm$^3$. A silicon photon-counting sensor (GMAPD) is used to detect the
scintillation light. 

Information about the working principle of the GMAPD can be found elsewhere (\emph{i.e.} \cite{mapd,mapd2}). In order to provide 
sufficient dynamic range of the detector, the so-called Deep-Microwell GMAPD (DM-GMAPD) with 1.5$\times$10$^5$ micro-pixels is used in our application.
The approach outlined in this paper allows the construction of ultra-compact hand-held  alpha-beta selective radiometers with the capability
of quantitative and qualitative analysis of the radioactive contaminants. Such instruments can be used in field environment, as well.
The design presented in this paper is part of a R\&D program on silicon photon-counting detectors
 that we have been conducting in Bern since a few years (see for instance \cite{Sebastien}).

\section{Detector design and test measurements}

\begin{figure}[htbp]	
\center\includegraphics[angle=0, width=0.5\textwidth]{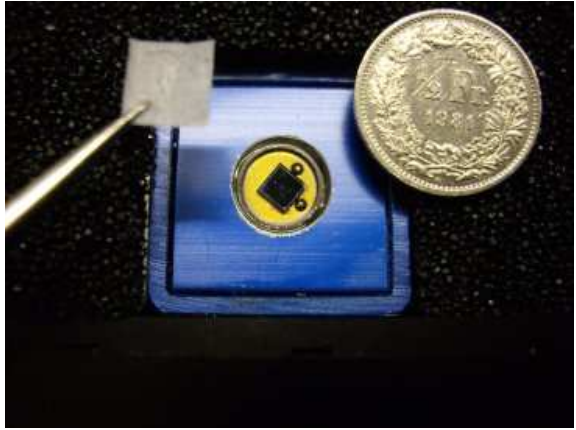}
\caption{G-MAPD detector front view. A 50 cent Swiss coin is shown for comparison. A sheet of filtering paper soaked in liquid scintillator is visible on the left side.}\label{detector}
\end{figure}

As photon-counting detector we used a DM-GMAPD  "MAPD-3N" from Zecotek\footnote{Zecotek Photonics Singapore Pte Ltd,  20, Science Park Rd, 01-23/25, Teletech Park, Science Park II, Singapore, 117674 }. This diode has a pixel density
of 15000 mm$^{-2}$ and a sufficiently high photon detection efficiency of the order of 30\% for green light (see \cite{mapd3}). The active area of the detector is a square of 3$\times$3~mm$^2$ (Figure~\ref{detector}).

The sample is a 5x5~mm sheet of 500~$\mu$m thick filtering paper
containing traces of radioactive isotopes (Figure~\ref{detector}, at the left). The isotopes were deposited onto the sample by wiping the surface  
of radioactive materials, such as $^{90}Sr$, $^{241}Am$ and $^{238}U$. The sample was soaked in the liquid scintillator and placed onto the detector window to provide an optical contact between the sensor and the sample.
A piece of aluminized Mylar is placed on top of the sample to improve light collection by reflection ($\approx$ 50\% increase).
The difference in refraction indices of the liquid scintillator (1.664) and the paper fibers (1.476) is rather low, so the wet sample is semi-transparent.
This creates good conditions for scintillation light collection and thus improves the energy resolution of the detector.
The sample sheet can be made of polymer fibers with a refraction index even closer to the one of the liquid, yet improving the light collection efficiency. For instance, polyethylene terephthalate fibers with the refraction index of 1.65-1.67 can be used for this purpose.

The thickness of the active scintillating layer is about 500~$\mu$m. The range of a 2 MeV alpha particle in the scintillator
is about 10 $\mu$m, so the full energy is absorbed in the active layer. However, the light yield is reduced by self-quenching, as described above. For beta particles of the same energy the range is about 10~mm. Thus,
only a fraction of the energy is deposited in the detector. These two distinct effects of light reduction bring the light yield from alpha and beta particles of the same energy to more or less the same range.

For our test we used a saturated solution of POPOP (1,4-Bis(5-phenyl-2-oxazolyl)benzene, Aldrich B50805) in 1-phenylnaphtalene (Aldrich P27402) as a non-extracting scintillator.
The use of specially developed high-yield extracting scintillators based on phosphoro-organic compounds 
will lead to a significant increase of the sensitivity and the
energy resolution. An even better capability of isotope identification can be achieved by using selectively
extracting compounds with improved PSD properties \cite{Alpha-Beta}. However, this is a topic for a dedicated R\&D and is out of scope of this paper.

\begin{figure}[htbp]	
\center\includegraphics[angle=0, width=0.7\textwidth]{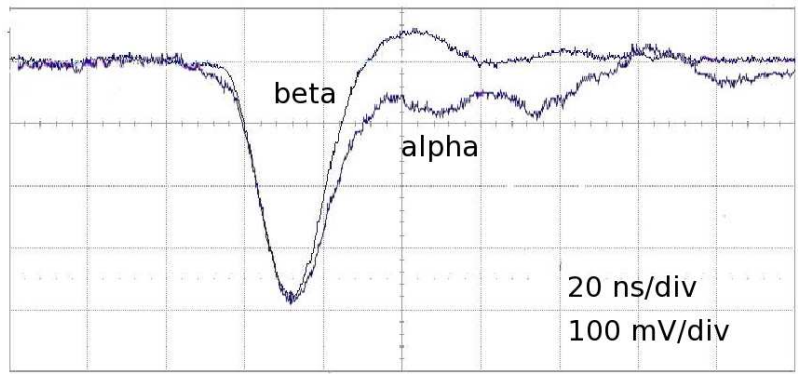}
\caption{Signal waveforms from scintillation produced by beta (electron) and alpha particles.}
\label{abwave}
\end{figure}

The signal from the diode was amplified with a fast low-noise shaping preamplifier
and supplied to the 
data acquisition electronics. There it was further amplified by a factor of 10 with a NIM UB/A5 Fast amplifier.
The difference between typical scintillation pulses from beta and from alpha particles can be seen in Figure \ref{abwave}.
The undershoot on the signal from beta particle is due to a parasitic pole in the frequency response of the amplification circuit. This pole has linear effect on the integrated signal from both beta and alpha particles, hence it does not affect the pulse shape discrimination.

\begin{figure}[htbp]	
\center\includegraphics[angle=0, width=0.7\textwidth]{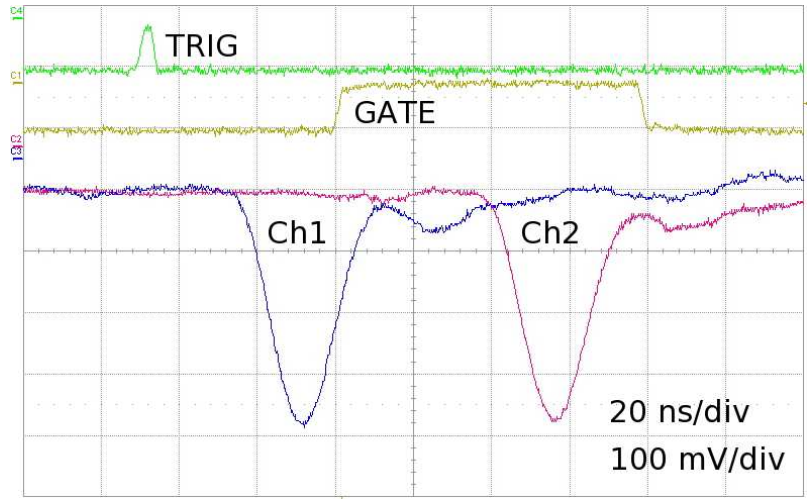}
\caption{Signal waveforms and timing. From the top: discriminator output pulse (TRIGGer), 
integration GATE pulse, two branches of the signal with different delays. The integration of the earlier (Ch1)
signal gives the slow component, while the integration of the later one (Ch2) provides the fast component.}
\label{a-wave}
\end{figure}

The signal was then split by an analog fan-out into two channels of a LeCroy A2249W fast integrating ADC (Ch1 and Ch2). The difference of the two channels is the delays with respect to the integration gate (80~ns long) as illustrated in Figure \ref{a-wave}. The ADC data are then read out by the computer using Ethernet CAMAC bus controller.

The channel with the earlier signal (Ch1) gives information about the slow component $S$ on the tail of the scintillation pulse, while the later one (Ch2) selects the fast component $F$ of the pulse. The ratio of these two components $R=S/F$ provides information about the density of the ionization produced by charged particles in the scintillator, which is used to discriminate alpha-particles from betas (electrons). 

The integration time is choosen to be relatively short for typical liquid scintillator PSD scenarios. This is due to a relatively high dark noise rate from GMAPD, so increasing integration time would reduce signal to noise ratio. However, choosing different integration times for fast and slow components could be a way to optimize the discrimination.

\begin{figure}[htbp]	
\includegraphics[angle=0, width=0.499\textwidth]{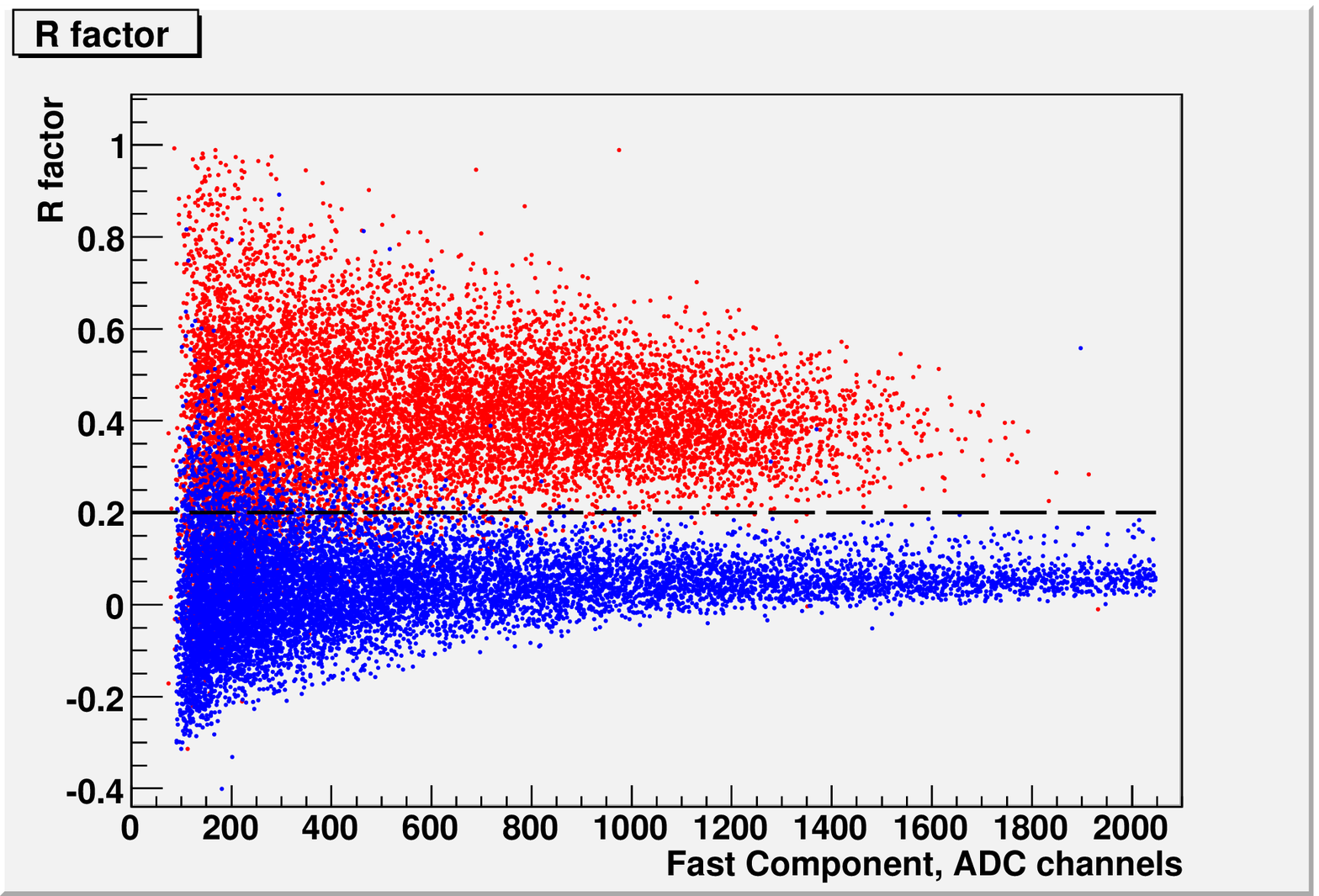}
\includegraphics[angle=0, width=0.499\textwidth]{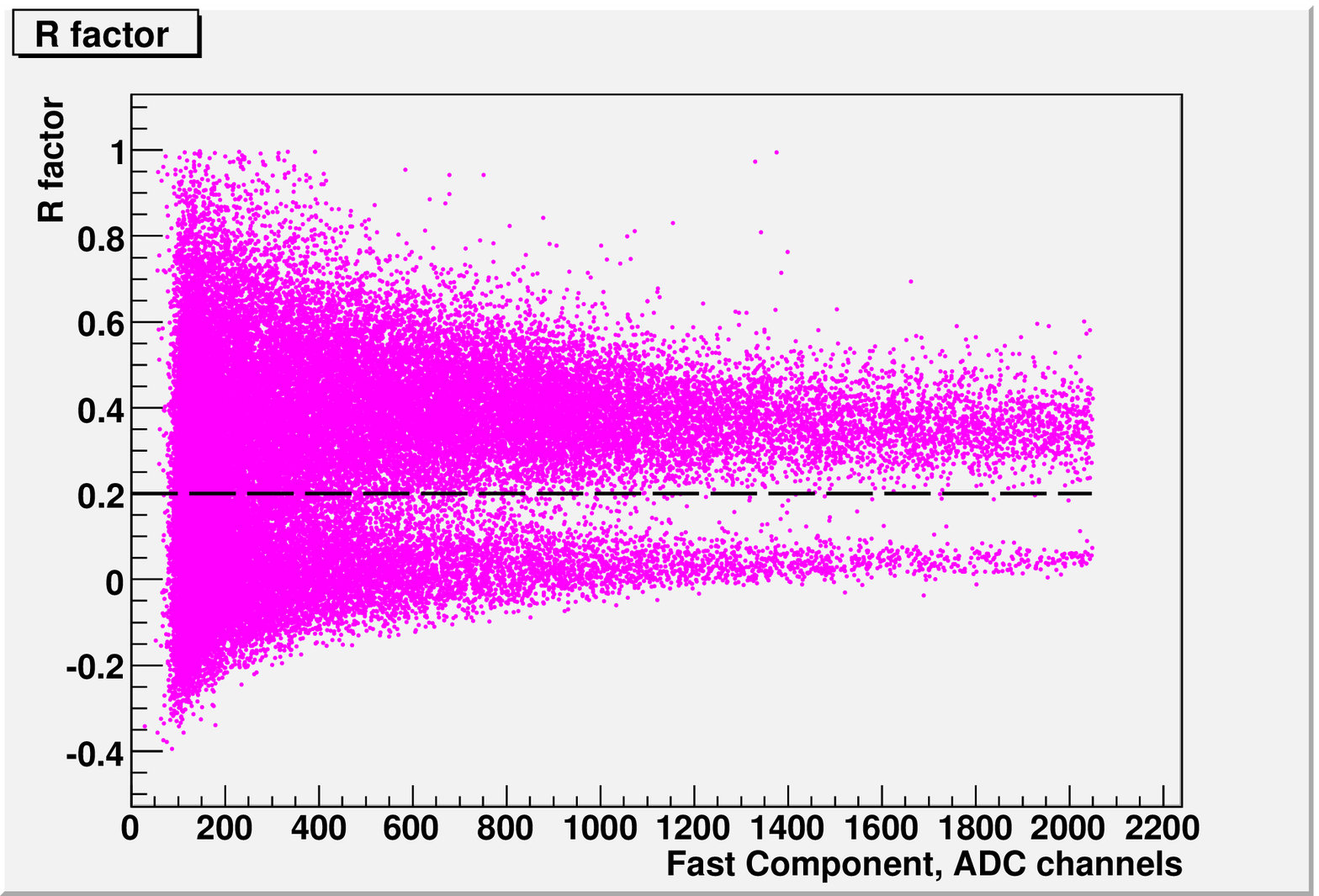}
\caption{R factor versus the fast component of the scintillation pulse for different radioactive sources.
Left panel: $^{241}Am$ alpha (top, red) and $^{90}Sr$ beta (bottom, blue) radioactive sources. 
The alpha-beta discriminating level is marked with a dashed line.
Right panel: $^{238}U$  radioactive source, alpha and beta activities are clearly visible on the two sides of the discriminating line at $R$=0.2.}
\label{U238}
\end{figure}

The left panel of Figure \ref{U238} shows a plot of $R$ versus $F$ signal amplitudes for $^{241}Am$ alpha (top, red) and $^{90}Sr$ beta (bottom, blue) radioactive sources. The alpha-beta discriminating level is depicted by the dashed line. On the right panel of Figure \ref{U238} the similar signal for natural Uranium (mostly $^{238}U$) is shown. This material has comparable alpha and beta activities. 

\begin{figure}[htbp]	
\center\includegraphics[angle=0, width=0.6\textwidth]{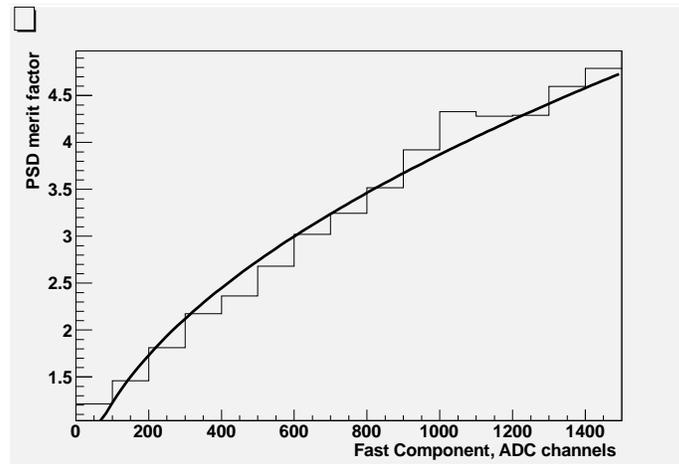}
\caption{The PSD merit factor D dependence on the scintillating light yield F fitted by $\sqrt{F}$ function.}
\label{Merit}
\end{figure}

\begin{figure}[htbp]	
\includegraphics[angle=0, width=0.499\textwidth]{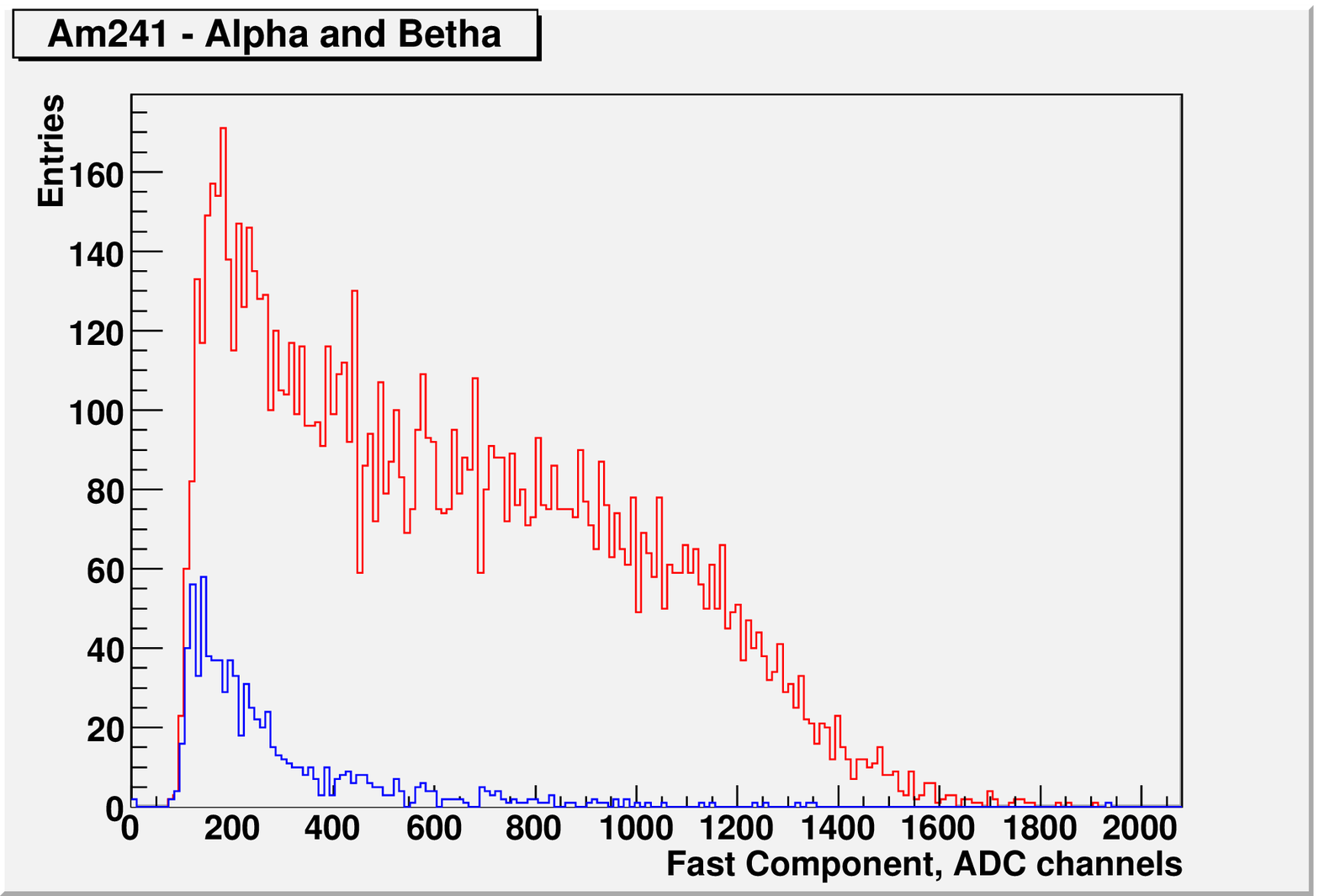}
\includegraphics[angle=0, width=0.499\textwidth]{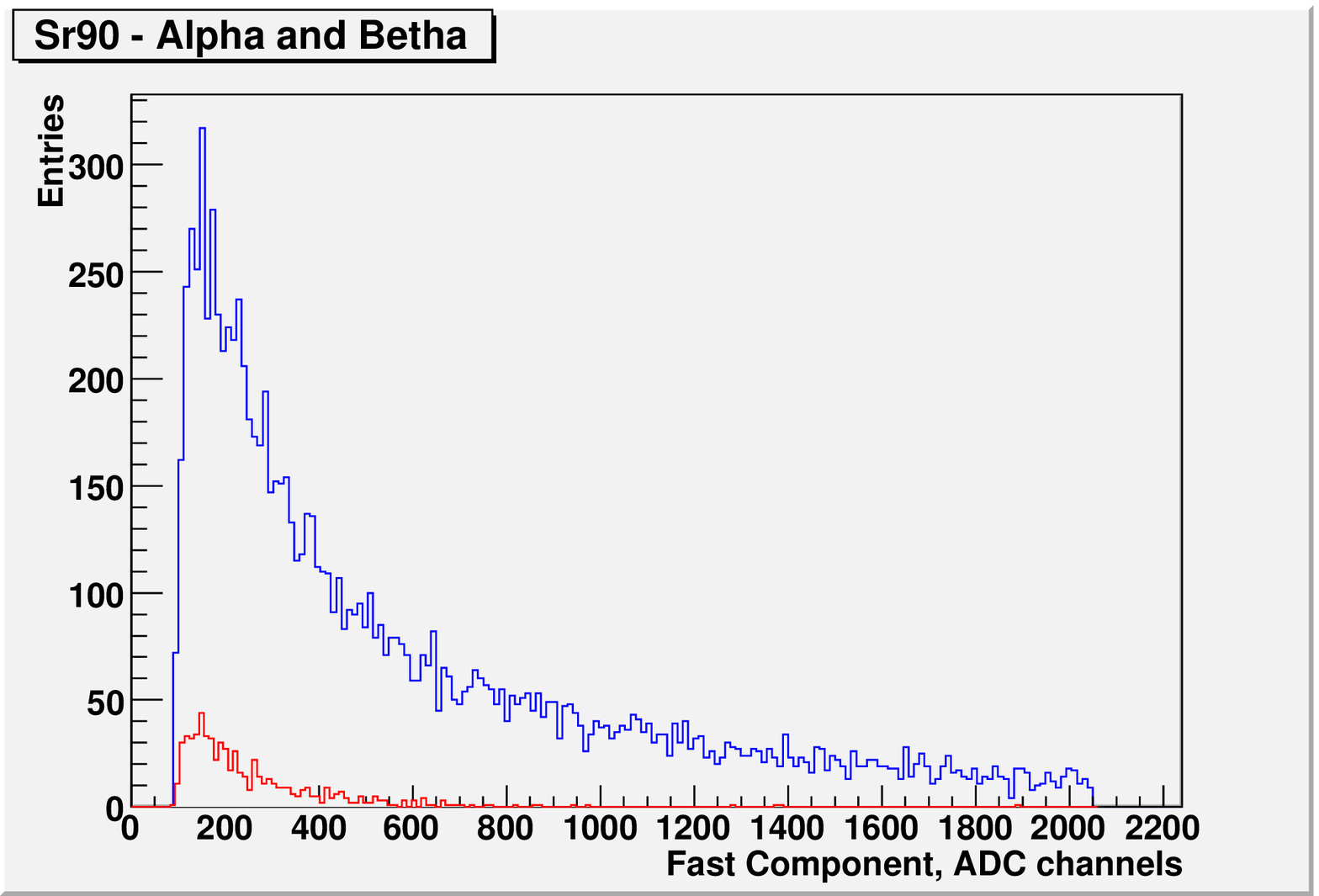}
\caption{The amplitude distributions of separated alpha-like (red) and beta-like (blue) components 
for  $^{241}Am$ (left) and $^{90}Sr$ (right).}
\label{SrAmSpectra}
\end{figure}

\begin{figure}[htbp]	
\center\includegraphics[angle=0, width=0.6\textwidth]{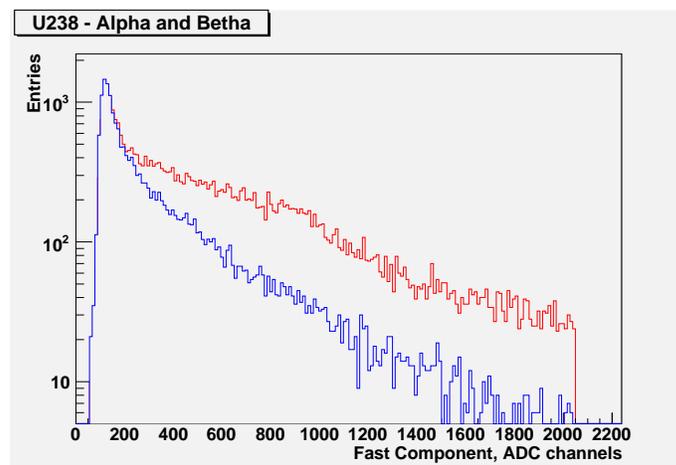}
\caption{The amplitude distributions of separated alpha-like (red) and beta-like (blue) components for natural Uranium.}
\label{USpectra}
\end{figure}

The efficient separation of signals from alpha and beta particles is clearly visible. The PSD separation capability is often quantified by the so called merit factor defined as

$$D=\Delta R/\sqrt{\sigma_\alpha^2+\sigma_\beta^2},$$ 

where $\Delta R$ is the difference in $R$, while $\sigma_\alpha$ and $\sigma_\beta$ are standard deviations of $R$ for alpha and beta particles. In general $\sigma_\alpha$ and $\sigma_\beta$ are proportional
to the square root of the number of detected photons. The merit factor is proportional to square root of the total light amount (Figure \ref{Merit}), so signals from alphas and betas start to overlap at low light, as seen in Figure \ref{U238}. The merit factor acheved is comparable to the one reported by the Borexino Collaboration, which used Pseudocumene as scintillating medium and conventional PMTs as photodetectors \cite{BorexinoPSD}. 

In Figure \ref{SrAmSpectra} one can see the amplitude distributions of separated alpha-like and beta-like components for $^{241}Am$ (left) and $^{90}Sr$ (right). The decay chain of $^{241}Am$ includes a fraction of low-energy photons and conversion electrons. The latter contribute to the beta-like component for this isotope. The light yield from 0.546~MeV beta of $^{90}Sr$ is similar to the one from 5.5~MeV alpha of $^{241}Am$ due to self-quenching. However, as already mentioned, due to the small volume and thickness of the scintillating sample, the beta response is significantly reduced.

In Figure \ref{USpectra} one can see the amplitude distributions of separated alpha and beta components for  $^{238}U$.
The decay chain of $^{238}U$ produces 6 beta and 8 alpha particles per nucleus. This ratio can not be seen over the full range of
scintillation amplitudes, since the separation at low energies vanishes. For high energies one clearly sees more alpha than beta.

The detector has a very low sensitivity to gamma radiation because of the low density and mass of its active volume (order of 1~mm$^3$ for a thickness about 500 $\mu$m, which corresponds to about $10^{-3}$ of a radiation length). 
This allows to overcome the usual difficulty of mixing beta and gamma activities in common PSD scintillating spectrometers.
The sensitivity of the detector to neutrons below 10 MeV is very low for the same reason (the detector thickness corresponds to $10^{-3}$ of the (n,p) scattering length).

\begin{figure}[htbp]	
\center\includegraphics[angle=0, width=0.6\textwidth]{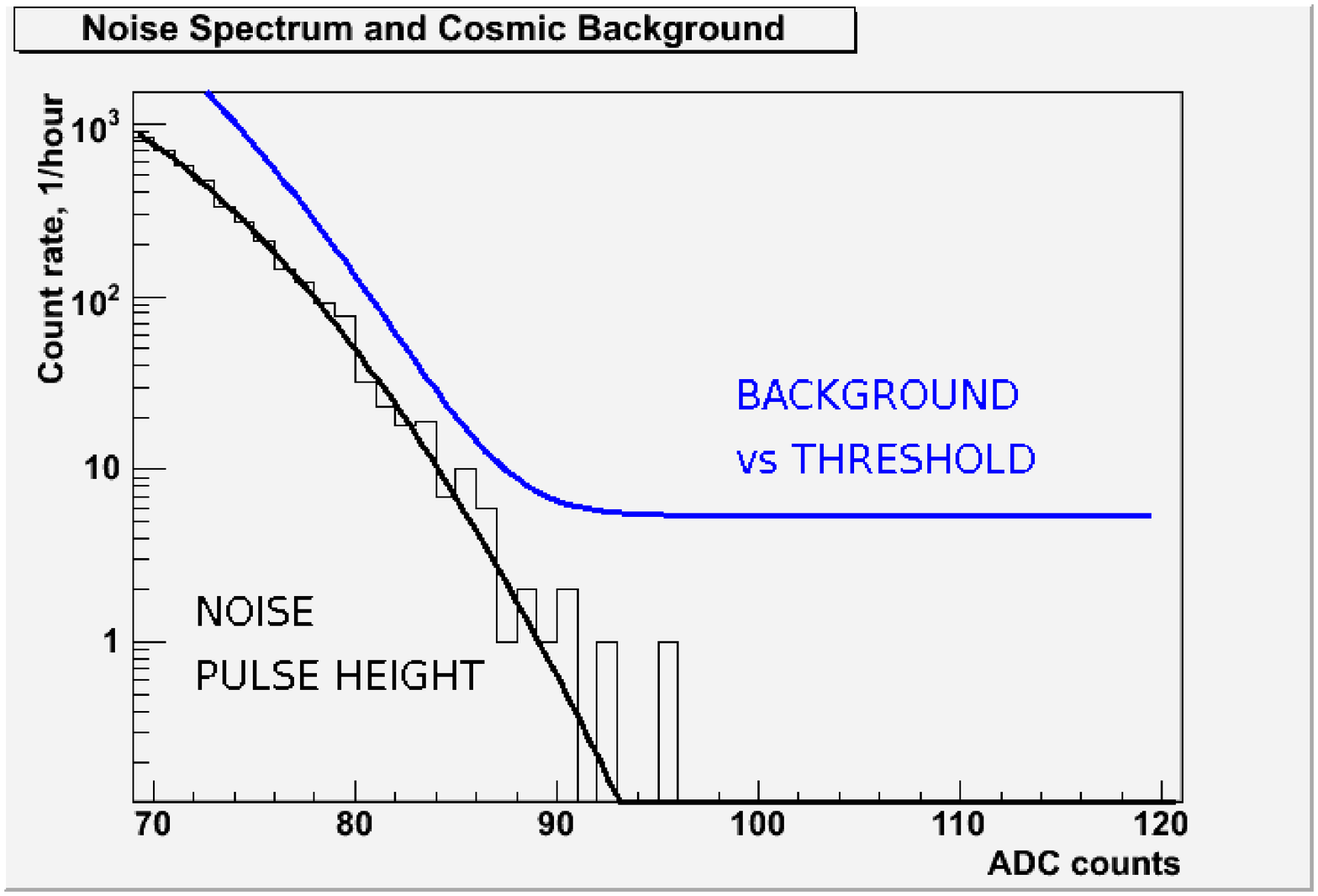}
\caption{Noise pulse height spectrum (black histogram and fitting curve) and noise rate versus signal threshold (blue curve). The flat part of the blue curve is due to cosmic radiation. Its level is calculated for a value of 1 particle/cm$^2$/min.}
\label{BG}
\end{figure}

The activity detection threshold of the proposed device is estimated by measuring the signal background without the sample.
Figure \ref{BG} shows the falling part of the pulse height spectrum of the noise signal. This spectrum is well fitted by a Gaussian (black curve). The integration of this signal from a given threshold to the range maximum (2048 channels) gives
the amount of noise counts versus the threshold value (blue curve). The flat part of this curve is due to cosmic-ray radiation, calculated for the known value of approximately 60 particle/cm$^2$/h (\cite{PDG}). Cosmic muons are mostly relativistic, so their response would be similar to beta, below the discriminating line in the R-factor plot.
The background activity in the beta-sector is about 1.7~mBq. The background activity in the alpha-like sector can be estimated
from the R-factor plot for $^{90}Sr$, which is a pure beta emitter. The fraction of beta-counts misidentified as alpha (R~>~0.2) is estimated to be less than 10\% for alpha energies above 0.3~MeV, which would set the higher limit to 0.17~mBq for the alpha-like background. 
The number of background counts after measurement time T[s] is given by 

$$N_{bg}[counts]=A_{noise}[counts/s]T[s],$$ where $A_{noise}$ is the noise pulse rate.

The RMS of this number defines the detection threshold for measured activity. Assuming a 1$\sigma$ separation from the noise, the minimum number of signal counts detected over the noise is  $N_{min}=\sqrt{N_{bg}}$. The minimum detectable activity is defined as 

$$A_{min}[Bq] = N_{min}/T = \sqrt{A_{noise}[counts/s]}/\sqrt{T[s]}.$$ 

This data gives an estimate of about 5/$\sqrt{T[s]}$~mBq for the beta activity and 1.7/$\sqrt{T[s]}$~mBq for the alpha activity detection thresholds.

\section{Conclusions}
An ultra-compact scintillating detector for applications in alpha-beta PSD spectrometry was developed and successfully tested.
This detector employs for the first time the advantages provided by Deep-Microwell Geiger-mode Multipixel Avalanche Photodiodes to detect scintillating light from micro-volumes of liquid scintillator probes. This allows to efficiently suppress 
cosmic-ray and ambient radiation background and to construct an ultra-compact device for field applications in monitoring radioactive contamination of various objects. There is a wide range of possibilities to further improve the performance of the spectrometer by using custom-designed high-yield extracting scintillating compounds. 

\section*{Acknowledgments}
We wish to thank our engineers and technical collaborators for their significant contributions to the presented work, in alphabethic order: J.~Christen, R.~H\"anni,  P.~Lutz, F.~Nydegger,  and H.-U.~Sch\"utz.

\end{document}